\documentclass[%
 reprint,
 amsmath,amssymb,
 aps,
prc,
]{revtex4-2}
\usepackage{braket}
\usepackage{graphicx}
\usepackage{hyperref}

\begin{document}


\title{Measurement of the Sequential \(3\alpha\) Process in the Photodissociation of \(^{12}\mathrm{C}\)}
\author{ Resmi K.Bharathan$^{1,2}$}
\email{resmikbharathan@gmail.com}
\author{Midhun C.V$^{1}$}
\author{ M.M Musthafa$^1$}
\email{mmm@uoc.ac.in}
\author{ Sreena M$^{1,3}$}
\author{Silpa Ajaykumar$^{1,2}$}
\author{Farhana Thesni M.P$^{1}$}
\author{Swapna B$^{1}$}
\author{Vafiya Thaslim T.T$^{1}$}
\author{Shaima A$^{1}$}
\author{Nived K$^{1}$}
\author{Akhil R$^{1}$}
\author{Anagha P.K$^{1}$}
\author{Arunima Dev T.V$^{1}$}
\author{Keerthi E.S$^{1}$}
\author{Akshay K. S$^{1}$}
\author{Arun P.V$^{1,3}$}
\author{S. Ghugre$^{4}$}
\affiliation{$^1$ Nuclear and Radiation Physics Group, Department of Physics, University of Calicut-673635,  India}
\affiliation{$^2$Medical Physics Group, Radiation Oncology, Malabar Cancer Centre, Thalassery-670103, India}
\affiliation{$^3$Department of Medical Physics, Indo-American Cancer Hospital \& Research Institute, Hyderabad-500034, India}
\affiliation{$^4$UGC-DAE-Consortium for Scientific Research, Kolkata, West Bengal 700098, India}
 
\date{\today}

\begin{abstract}
The cross sections for the $^{12}\mathrm{C}(\gamma,\alpha)^{8}\mathrm{Be}\rightarrow 3\alpha$ reaction have been successfully measured using exclusive coincidence between three $\alpha$ particles, minimizing Compton background. Sequential breakup kinematics are evident, and the cross sections are presented as locally averaged histogram values. Theoretical \textsc{Fresco} CDCC-CRC calculations reproduce the experimental data, showing that the process involves electromagnetic coupling to both $^{8}\mathrm{Be}^{0^+}$ and $^{8}\mathrm{Be}^{2^+}$ states. This study confirms that the $^{12}\mathrm{C}(\gamma,\alpha)^{8}\mathrm{Be}\rightarrow 3\alpha$ reaction proceeds via a sequential mechanism, crucial for understanding its significance in radiotherapy dosimetry.

\end{abstract}

\maketitle


\section{ Introduction}
High-energy photon beams, reaching up to 18 MeV, find widespread application in radiotherapy, where they deposit energy primarily through secondary electron interactions, quantified by absorbed dose. To effectively treat deep-seated tumors, higher energy bremsstrahlung beams are employed, but concerns arise from nuclear reactions initiated by these beams, particularly $^{12}$C$(\gamma,3\alpha)$, which generates high-energy $\alpha$ particles.\citep{kirsebom2010breakup}. These particles, when deposited within a few cubic micrometers, create micro hotspots, causing complete tissue destruction particularly in the target. However, current dosimetry techniques struggle to measure these events due to their small volume and the volume averaging effect of detector.\citep{rijin2022effect} This issue is critical given the abundance of $^{12}$C atoms in biological tissue and the high photon flux used in radiotherapy. The clinical significance of micro hotspots is determined by their mean separation and behavior, which are influenced by the $^{12}$C($\gamma$, 3$\alpha$) cross section and the alpha spectrum, respectively. Understanding these parameters is crucial for optimizing radiotherapy treatments.
\par Radiation transport simulations are increasingly favoured over traditional dosimetric measurements for the prediction of micro hotspots in radiotherapy, primarily due to inherent limitations in the latter approach. These simulations necessitate a meticulous examination and analysis of the $^{12}$C($\gamma$, 3$\alpha$) reaction cross sections, as they play a fundamental role in the accurate depiction of radiation interaction dynamics. However, despite advancements, the full comprehension of the $^{12}$C($\gamma$, 3$\alpha$) reaction kinematics remains elusive, particularly in scenarios where the excitation energy of $^{12}$C surpasses the critical threshold of the $\alpha$ separation energy of 7.36 MeV. Above this energy threshold, the behavior of the reaction and the ensuing particle spectrum become increasingly complex under three body kinematics, with higher excitation levels of $^{12}$C resulting in a proportional increase in the kinetic energy exhibited by the resultant $\alpha$ particles upon their breakup. Moreover, the presence of resonant couplings within the exit channel of the reaction further complicates the determination of the excitation function and particle spectrum, introducing significant variability and unpredictability into the overall process. 
\par The $^{12}$C($\gamma$,3$\alpha$) reaction is believed to follow a sequential process, beginning with the capture of a photon by $^{12}$C, which then leads to the population of the $^8$Be+$\alpha$ state via electromagnetic coupling. Subsequently, due to the unbound nature of $^8$Be, it undergoes instantaneous breakup into two $\alpha$ particles. Additionally, resonant states of $^8$Be contribute to the reaction, alongside overlaps from the $^8$Be$\rightarrow \alpha + \alpha$ breakup continuum. These various couplings and overlaps are expected to alter both the $^{12}$C($\gamma$,$\alpha$)$^8$Be$\rightarrow$3$\alpha$ excitation function and the resulting $\alpha$ spectrum. Hence, besides the excitation function, the resonant couplings has to be well understood for generating evaluated data sets to be applied in simulations.


\par There are numerous measurements on $^{12}$C, break up into $3\alpha$ reaction. However, most of them are limited to the resonant breakups of $^{12}$C, which are different from the electromagnetic coupling. Kirsebom et al.\citep{kirsebom2009observation, PhysRevCKirsebom} has studied the sequential breakup of $^{12}$C$\rightarrow ^8$Be$^* + \alpha \rightarrow 3 \alpha $ via the $^{10}$Be$(^3\mathrm{He},p)^{12}$C reaction and identified an electromagnetic coupling to $^8$Be$^* + \alpha$ states. However the measurement is with a limited statistics. Similar kind of electromagnetic coupling has been observed by Laursen et al.\citep{laursen2016unbound} in the $^{11}\mathrm{B}+p$ reaction, decaying to $^8$Be$^* + \alpha$ states. Kirsebom et al.\citep{kirsebom2020experimental} has identified the sequential breakups of unbound resonant states $(2^-,1^+ \: \mathrm{and} \: 4^-)$ of $^{12}$C via the resonant states of $^8$Be. Similarly, weak electromagnetic couplings in the transition from Hoyle state has been observed by T.K Rana et al \citep{RANAPhysLetB, RanaPRC}. These works are based on the population of $^{12}$C states via proton capture, transfer induced or $\beta$ delayed pathways. These reactions are populating the resonant states of $^{12}$C than the $^8$Be$* + \alpha$ continuum, hence these results cannot directly adapted for $^{12}$C($\gamma$,$\alpha$)$^8$Be$\rightarrow$3$\alpha$ reaction phase.
\par There are several challenges existing to execute the measurement of $^{12}$C($\gamma$,$\alpha$)$^8$Be$\rightarrow$3$\alpha$ reaction such as the availability of monoenergetic $\gamma$ for sufficient intensity, isolating Compton electrons and  other scattered photons eventing on the detectors. Since the bremsstrahlung beams of higher intensities are available, reconstructing true events from the higher background level is the major concern.

\par Hence to address the real situation of populating $\alpha + ^8$Be (ground and $0^+$) through electromagnetic coupling, an experiment on $^{12}$C($\gamma$,$\alpha$)$^8$Be$\rightarrow$3$\alpha$ has been performed, and  successfully detected the 3$\alpha$ particles, with exclusive event identification, by minimizing the Compton background. The measured cross sections has been reproduced with {\sc fresco} CDCC-CRC calculations.
\begin{figure}
    \centering
    \includegraphics[width=\columnwidth]{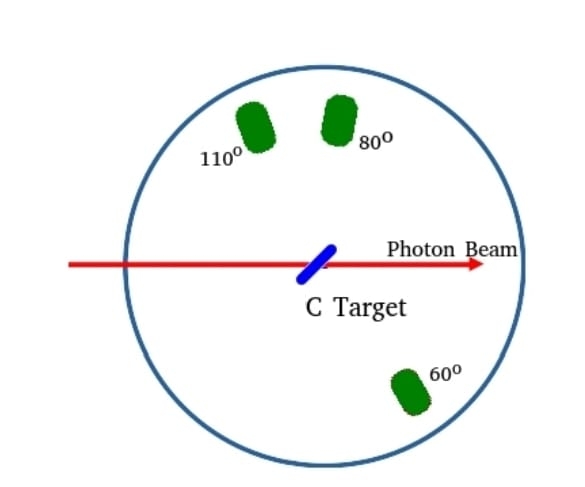}
    \caption{Schematic diagram of the experimental set up showing the Silicon detectors positions and target orientations}
    \label{fig:1}
\end{figure}

\section{\label{sec:level2}Materials and Methods}
\begin{figure}
    \centering
   \includegraphics[width=\columnwidth]{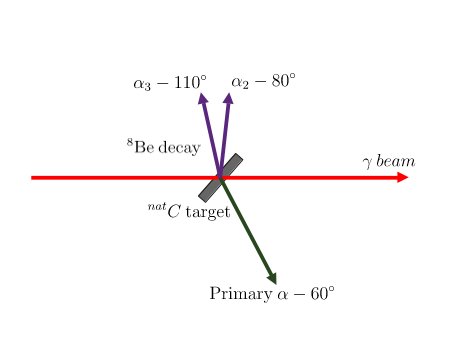}
    \caption{Calculated event kinematics for achieving the exclusive coincidence}
    \label{fig:3}
\end{figure}

The reaction, $^{12}$C($\gamma$,$\alpha$)$^8$Be$\rightarrow$3$\alpha$, has been performed with 14.6 MeV end point energy bremsstrahlung beam, as there is no mono energetic source exists in the current scenario. The experimental setup has been designed in such a way to detect the $3 \: \alpha$s in coincidence, for a limited solid angle, which made the capability of reconstruction of the photon energy associated to each individual event. The detector angles are chosen as per the three body sequential kinematics given by R. J. de Meijer and R. Kamermans \citep{RevModPhys.57.147}. The schematic diagram of experimental setup is given in Figure \ref{fig:1}. The illustration of kinematically correlated three body events are shown in Figure \ref{fig:3}.
\\
\par
\begin{figure}
    \centering
    \includegraphics[width=\columnwidth]{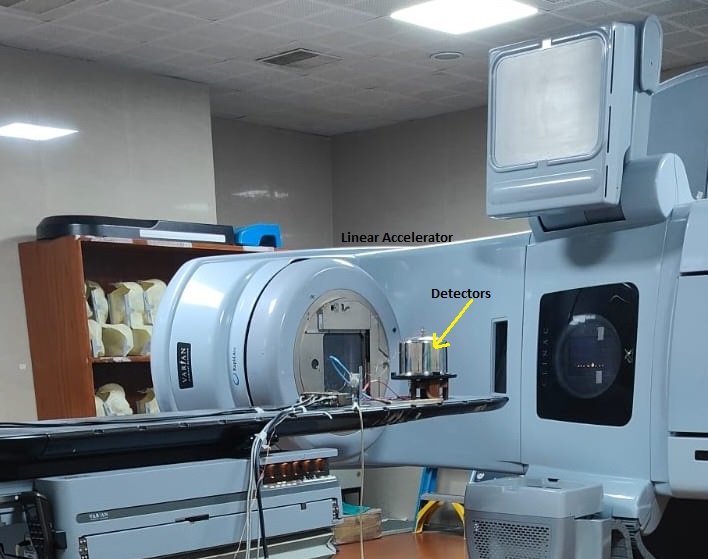}    
    \caption{Experimental setup showing Clinac-iX Medical Linear Accelerator, with scattering chamber on the treatment couch}
    \label{fig:2}
\end{figure}

 The primary $\alpha$ angle is set at $+60^\circ$ to mitigate the significant influence of Compton electrons and photons at extreme forward angles. To achieve exclusive coincidences, other detector angles are meticulously chosen at $-80^\circ$ and $-110^\circ$, as determined by relativistic kinematics calculations. The experimental setup, depicted in Figure\ref{fig:2}, illustrates this configuration. Conducted at the Varian Clinac-iX accelerator facility at Malabar Cancer Centre, Thalassery, the experiment utilized a bremsstrahlung beam with an endpoint energy of 14.6 MeV. This beam, meticulously collimated, was directed towards a 300 $\mu$m thick carbon target ($^{nat}$C). The real experimental setup with Varian Clinac-iX and the scattering chamber mounted on the treatment couch is shown in Figure \ref{fig:2}. 

\par Three silicon surface barrier detectors, each 1500 $\mu$m thick, were employed to detect the $\alpha$ particles. The detector collimators are chosen as aluminum, to minimize the electrons generated from the collimator, reaching the detectors. These detector signals were preamplified using Ortec-IH142 preamplifiers, and the data was acquired via a CAEN- DT5743 digitizer. These detectors were biased at an optimal voltage to ensure a well-measurable pulse rise time. To minimize pulse height aberrations at reduced voltage, the energy was mapped into charge rather than peak voltage. The detectors were calibrated using 5.48MeV $\alpha$ particles from $^{241}\mathrm{Am}$ source. A precision pulser has been configured with 100Hz, to ensure the calibration, by identifying the linearity.  Particle identification was achieved through a 2D correlation between charge and pulse rise time, similar to the method described in \citep{PSD}. However with pulse rise time instead of zero crossover timing. Additionally, proper CFD backings were configured for the detectors to prevent pileups caused by Compton electrons. In addition to recording individual detector energies, a multiplicity trigger and energy sum generated from CFD were also acquired to enable fast identification and scaling of exclusive three-body coincidences. The Compton events were figured with PSD and rejected from event registration in the offline analysis of the waveform. The  measurement has been performed with the external trigger mode, with the multiplicity $\ge \: 2$.

\par To determine the flux of the photon beam, accelerator-built transmission-type beam monitor chambers were employed, maintaining a consistent  flux of approximately $10^{13}\: \mathrm{photons}/cm^2 $ throughout the experiment. The built-in monitor chambers were calibrated prior to the experiment using secondary standard ionization chamber, to a calibration of 1 monitor unit (MU) equals 1 cGy. This is converted into flux by Kerma formalism used for dosimetry\citep{kerma}.

\par
\begin{figure}
    \centering
    \includegraphics[width=\columnwidth]{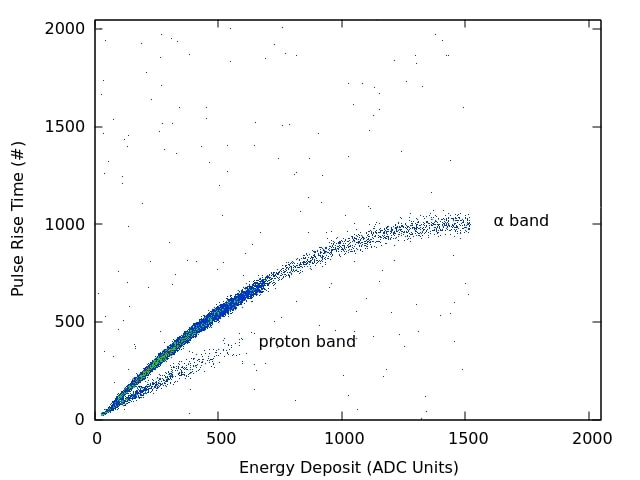}
    \caption{The Charge Pulse shape correlation showing the particle discrimination}
    \label{fig:4}
\end{figure}

\begin{figure}
    \centering
    \includegraphics[width=\columnwidth]{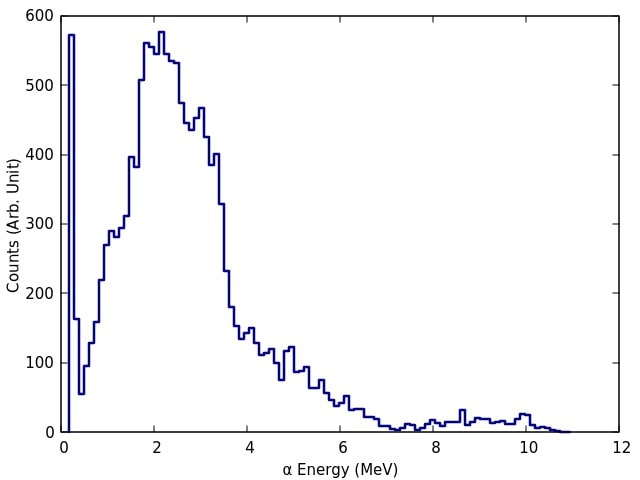}
    \caption{The measured row $\alpha$ spectrum at 60$^\circ$ detector}
    \label{AlphaSpectrum}
\end{figure}

 A 2D plot between charge and rise time has been constructed for each detector. The typical particle discrimination plot is shown in Figure \ref{fig:4}. The proton and $\alpha$ bands are well evident without overlaps in the present configuration. The $\alpha$ events of each detector were banana gated for event localization.  A total energy parameter, gated for triple coincidence alpha events has been generated. The $\alpha$ spectrum identified in the $60^\circ$, without triple coincidence gate, is given in Figure \ref{AlphaSpectrum}. This has been summed with the breakup Q value for obtaining the photon energy subjected for the break up. It is assumed that the 92 keV added by the $^8$Be decay is lesser than the binning resolution, hence it is excluded.  

\begin{figure}
    \centering
    \includegraphics[width=\columnwidth]{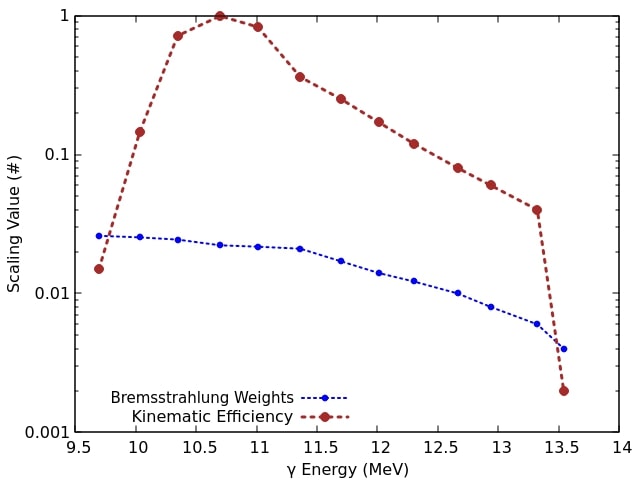}
    \caption{Kinematic efficiencies and Bremsstrauhlung yield for computing each excitation energy }
    \label{fig:5}
\end{figure}

The events in the photon energy domain has been binned into locally averaged histogram values. The event yields in each energy bin has been converted into cross sections by accounting the photon flux corresponding to each bremsstrahlung energy bin and the three body kinematic efficiencies. The intensity scaling factors were accounted from the segmental integration of the bremsstrahlung spectrum. The spectum for the 14.6 MeV bremsstrahlung has been adapted through the method by Midhun et al.\citep{midhun2020spectroscopy}. Further the kinematic efficiencies for the experimental setup has been estimated through Monte-Carlo approach on 3 body sequential emission kinematics. The calculated kinematic efficiencies and bremsstrahlung weights are given in Figure \ref{fig:5}. The obtained cross sections were compared with {\sc fresco } CDCC-CRC calculations\citep{THOMPSON1988167}.
\section{{\sc fresco} Calculations}
The concept adapted for theoretical calculations is structured such that the $\gamma + ^{12}\mathrm{C}$ entrance channel state couples to the final $\alpha + ^8\mathrm{Be}$ states through electromagnetic modes. For this current analysis, the $E_1$ mode is considered. As the final state forms, the ejectile $\alpha$ wave function couples to the $^{8}\mathrm{Be}$ resonant or breakup continuum, resulting in a sequential breakup. This concept has been implemented through \textsc{Fresco} CDCC-CRC calculations.

\par The entrance channel mass partition $^{12}$C+$\gamma$ has been defined with a pseudo potential for photons. $\alpha$+$^8$Be has been defined as the exit channel. The $^8$Be is binned into $\alpha$ core and $\alpha$ valance structure. Potential having Gaussian shape, with $V_0 = 106\: MeV$ and $R_0 = 2.236 \: fm$, is used to define the $\alpha$+$\alpha$ breakup continuum state of $^8$Be. The Guassian potential parameters has been retrieved from Marquez et al.\citep{Marquez} and Rusek et al.\citep{Rusek}. The $0^+ \: \mathrm{(ground)}$ and $2^+ \: \mathrm{(3.03 \: MeV)}$ states of $^8$Be has been considered. These states has been defined with $p+^7Li$ binning potential, having Woods-Saxon volume form with parameters $V_0 =50.6\: MeV$, $r_0 = 1.15 \:fm$ , $a_0 = 0.57 \:fm$, $V_{so} =5.5\: MeV$, $r_{so} =1.15\: fm$ and $a_{so} =0.57\: fm$. The potential shape and parameters were adapted from Chattopadhyay et al. \citep{chattopadhyay2018resonant}.
\par Along with the CRC states, the break up continuum generated by $\braket{\alpha+\alpha}{^8\mathrm{Be}}$ is also considered. The breakup continuum generated through the relative movement of $\alpha - \alpha$ cluster in $^8$Be, has been discritized based on the wave number and angular momentum, based on,
\begin{equation}
    k=\frac{\sqrt{2 \mu E_x}}{\hbar^2}
\end{equation}
Where $\mu$ is the reduced mass of $\alpha - \alpha$ system, and $E_x$ being the excitation energy. The discritized states of continuum having bin width of 0.175 fm has been generated upto $k^{-1} = 0.7\: fm $. The spin of each bin has been taken as the relative angular momentum of the $\alpha - \alpha$ system at that angular momentum where as the spin of $\alpha$ is zero by default. $3\alpha$ folding potentials with Gaussian form factor, with parameters described above, has been used for defining these states. An overlap of these CDCC states over the $0^+$ and $2^+$ CRC states has been defined, for accounting the width broadening of the resonant states. 
\par All the states in exit channel mass partition has been made coupled to the entrance channel mass partition electromagnetic coupling of $E_1$ mode (electric dipole mode). The spectroscopy factors for $0^+$ and $2^+$ CRC states has been used as unity, adapting the value from D.Chattopadhyay et al.\citep{chattopadhyay2018resonant}. An average spectroscopy factor of 0.75 has been used for CDCC states. However it couldn't be verified in the current scenario as due to the systematic uncertainties and poor discrimination in the relative energy determination.  The {\sc fresco} has been run for CRC only for the first instance and the CDCC+CRC  for the second instance.  The cross section for each state has been taken and 60$^\circ$ cross sections has been selected.  Each bin energy has been converted to $\gamma$ beam incident energy using relativistic two body kinematics.
The contributions from individual states has been added to get the final cross sections.  The final cross sections has been compared with experimental cross sections.
 
\section{\label{sec:level3}Result and Discussion}
\begin{figure}
    \centering
    \includegraphics[width=\columnwidth]{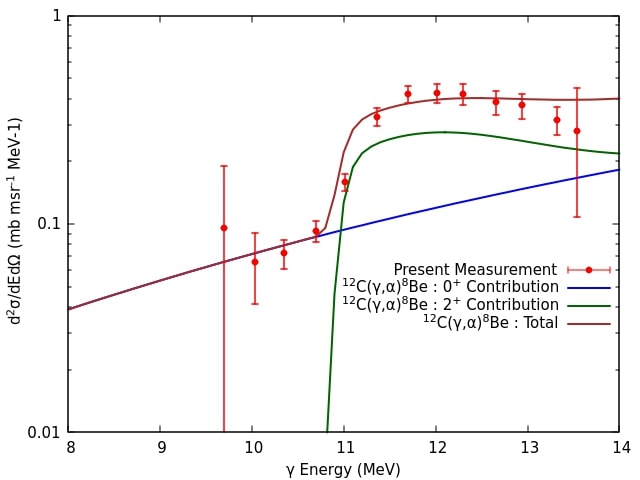}
    \caption{Experimental cross section adding with FRESCO calculation for $0^+$ and $2^+$ contributions}
    \label{fig:6}
\end{figure}

The $^{12}$C($\gamma$,$\alpha$)$^8$Be$\rightarrow$3$\alpha$ cross sections has been successfully measured by constructing exclusive coincidence between 3$\alpha$ particle events, by establishing multiplicity trigger condition. The events were recorded with minimum Compton background, unlike the previous attempts. The events under sequential kinematics is illustrated in Figure \ref{fig:4}.  The cross sections are presented as locally averaged histogram values, in the fraction of photon energy, in Figure \ref{fig:6}.  The present cross sections belongs to inclusive cross section of all states of $^8$Be, populated in the present scenario, as the relative energy resolution provided by the present method is not enough to identify the individual breakup contributions.  Since the geometrical efficiency of the present set up by the 3 body sequential emission kinematics is similar to the total efficiency. Hence the same has been used.  This adds up a systematic uncertainty of 1.3\% to the cross section data.
\par The {\sc fresco} CDCC – CRC calculations are shown as solid lines in the Figure \ref{fig:6}. The blue line show the contributions from  $^{12}$C($\gamma$,$\alpha$)$^8$Be$\rightarrow 3 \alpha$,  corresponding to the $\alpha$ production from the ground state of $^8$Be, and Green curve shows the $\alpha$ production from the  $2^+$ resonance state of $^8$Be (3.03 MeV). The coherent sum of the contributions from $0^+$ and $2^+$ state of $^8$Be explains the experimental cross section. Based on the experimental measurement, with the support of the theoretical calculations, it can elucidate that the $^{12}$C($\gamma$,$\alpha$)$^8$Be$\rightarrow 3\alpha$  process is a sequential process.
\par The $\gamma+^{12}$C initial capture state populates $\alpha+^8$Be state through electromagnetic coupling in direct mode to $^8\mathrm{Be}^{0^+} +\alpha $ or $^8\mathrm{Be}^{2^+} +\alpha$ states. The present measurement followed by the CDCC-CRC identifies the presence of the coupling of $2^+$ state in the excitation function, above the $E_\gamma$ of 10.8 MeV, where as the Q-value for populating this state is 10.39 MeV. The $2^+$ component is showing a broad resonant width due to the overlap of $\braket{\alpha+\alpha| ^8\mathrm{Be}}$ breakup continuum.
\par Identification of the electromagnetic couplings to $^8\mathrm{Be}^{0^+} +\alpha $ and $^8\mathrm{Be}^{2^+} +\alpha$ states can be coincided along with the observation by Laursen et al.\citep{laursen2016unbound} and Kirsebom et al. \citep{PhysRevCKirsebom, kirsebom2020experimental}.  There Kirsebom et al. \citep{PhysRevCKirsebom} discusses the population of the $2^+$ state of $^8$Be through the decay of 12.71 MeV $(1^+)$ state of $^{12}$C. However it was holding an ambiguity about the sequential breakup mechanism as the life time of the $2^+$ state of $^8$Be is about $1.5 \: MeV$. The present study based on the measurement with direct photons, followed by CDCC-CRC analysis shows that the population of $^8$Be states are due to the direct electromagnetic couplings. However, the continuum states are existing as a equal counter part. By considering the $^7\mathrm{Li}+p$ binning potential for these states, the the population has been well explained and the ambiguity is thereby clarified. 

\par The exclusive cross sections are  typically in fractions of millibarns, and the average energy of the $\alpha$ particles observed is around 3 MeV. Considering these cross sections, photon flux used in radiotherapy, and the number of $^{12}$C nuclei present in the unit tissue, the $^{12}\mathrm{C}(\gamma,\alpha)^8\mathrm{Be} \rightarrow 3\alpha$ events are very significant, as each event deposit almost 170 Gy dose in a 10$\mu m^3$ volume. As considering serial organs like spine and other neutral tissue, this makes a permanent damage, and may leads to a paralysis of the serial organ. Further extensive measurements with greater solid angle coverage and higher statistics are required to improve our understanding of this process.

\section{Conclusion}
The sequential breakup mechanism of $^{12}\mathrm{C}(\gamma,\alpha)^8\mathrm{Be} \rightarrow 3\alpha$ has been experimentally observed in the $\gamma$ energy range of 9.5 MeV to 13.5 MeV. The experimental cross sections were successfully reproduced using {\sc fresco} CDCC-CRC calculations. These calculations indicate that the process is sequential, the electromagnetic coupling directly populates $^8\mathrm{Be}^{0^+} +\alpha $ and $^8\mathrm{Be}^{2^+} +\alpha$ states. The overlap of breakup continuum on the resonant states makes the resonant widths broader.
\section{Acknowledgements}
The authors gratefully acknowledge the support of Dr. Geetha Satheeshan, Head of the Department of Radiation Oncology at Malabar Cancer Center, Thalassery, as well as the invaluable assistance provided by the radiotherapy technical staff.
\par The authors extend their sincere thanks to UGC-DAE-CSR-KC for providing financial support for this project. 

\bibliographystyle{apsrev4-2}
\bibliography{12CPhotodis}

\end{document}